\input{aipcheck}

\documentclass[
    ,final            
  ]
  {aipproc}

\layoutstyle{8x11single}

\begin{document}

\title{One-pion production in neutrino-nucleus collisions}

\classification{13.15.+g,25.30.Pt}
\keywords{One-pion production by neutrinos}

\author{E. Hern\'andez}{
  address={Departamento
de F\'\i sica Fundamental e IUFFyM,\\ Universidad de Salamanca, E-37008
Salamanca, Spain.},
  email={gajatee@usal.es},
}

\author{J. Nieves}{
  address={Instituto de F\'\i sica Corpuscular (IFIC), Centro Mixto
  CSIC-Universidad de Valencia, Institutos de Investigaci\'on de
  Paterna, Apartado 22085, E-46071 Valencia, Spain},
  email={jmnieves@ific.uv.es},
}
\author{M.J. Vicente-Vacas}{
  address={Departamento de F\'\i sica Te\'orica e IFIC, Centro Mixto
  Universidad de Valencia-CSIC, Institutos de Investigaci\'on de
  Paterna, Apartado 22085, E-46071 Valencia, Spain},
  email={Manuel.J.Vicente@uv.es},
}


\begin{abstract}
We use our model for neutrino pion production on the nucleon to study 
pion  production on a nucleus. The model is conveniently modified to
include in-medium corrections and its validity is extended up to 2\,GeV neutrino
energies by the inclusion of new resonant contributions in the production
process.
Our results  are compared  with recent
MiniBooNE data measured in mineral oil. Our total cross sections are 
below data for neutrino energies above $\approx 1\,$GeV. As with other
theoretical calculations, the agreement with data
 improves if we neglect
pion final state interaction. This is also the case for differential cross
sections convoluted over the neutrino flux.

\end{abstract}


\maketitle
\section{Introduction}
The MiniBooNE Collaboration has recently published one pion production
cross sections on mineral oil ($CH_2$) by $\nu_\mu/\bar\nu_\mu$ neutrinos with energies
below 2\,GeV 
~\cite{AguilarArevalo:2010bm,AguilarArevalo:2010xt,AguilarArevalo:2009ww}. 
These are the first pion
production cross sections to be measured since the old bubble chamber experiments
carried out at  Argonne National Laboratory (ANL)~\cite{Campbell:1973wg,Radecky:1981fn} and
Brookhaven National Laboratory (BNL)~\cite{Kitagaki:1986ct}. The latter were
measured in deuterium where nuclear effects are 
small~\cite{AlvarezRuso:1998hi,Hernandez:2010bx}. 
MiniBooNE data
 poses an extra problem to theoretical models  due to the expected relevance of
in-medium modifications  and final state 
interaction (FSI) in carbon. In Ref.~\cite{Golan:2012wx}, the use 
of a formation time/zone, that  reduces the impact of FSI, 
leads to a good agreement with the shape of different neutral current ($NC$)
$\pi^0$
production differential cross
sections. Charged current ($CC$) single pion
production off $^{12}C$ for neutrino energies up to 1\,GeV is analyzed in
Ref.~\cite{Sobczyk.:2012zj}. Their results for
 total cross sections are below
MiniBooNE data in the high neutrino energy region ($0.8-1\,$GeV) and the agreement
improves if FSI is neglected. A different approach valid only in the low $Q^2$ region
is presented in Ref.~\cite{Paschos:2012tr}. There the authors evaluate pion
production by neutrinos in the low $Q^2$ region and for neutrinos 
in the energy range $0.5\sim 2$\,GeV. The model is based on partial
conservation of the axial current (PCAC) hypothesis, the conserved vector current
(CVC) hypothesis and the use of experimental cross section data at the nucleon
level. The agreement found with MiniBooNE
data is good for $Q^2$ values up to 0.2\,GeV$^2$. In 
Ref.~\cite{Lalakulich:2012cj} the authors use 
  the Giessen Boltzmann-Uehling-Uhlenbeck (GiBUU) model  finding  that 
 total cross sections measured by MiniBooNE are higher than 
 theoretical ones for neutrino energies above $0.8\sim0.9$\,GeV. As in 
 Ref.~\cite{Sobczyk.:2012zj}, the agreement with data for total and different 
 flux
 averaged differential cross sections
is better if pion FSI is neglected. However, 
 as also shown in  Ref.~\cite{Lalakulich:2012cj},
the same FSI model applied to pion photoproduction  is able to give a fair
reproduction of experiment in that case.

In this contribution we  address  the problem of pion production in a nucleus
starting from  our pion
production model at the nucleon level taken from 
Refs.~\cite{Hernandez:2007qq,Hernandez:2010bx}. In order  to better
compare to MiniBooNE data we extend the model up to
2\,GeV neutrino energies, well above the $\Delta$ resonance region for which it
 was originally developed. Above the Delta region also the $D_{13}(1520)$ 
 resonance plays a role
 ~\cite{Leitner:2008ue} and in the present calculation we include its 
 contribution.  We  also take into
 account in-medium corrections to the production process. Those are Pauli-blocking 
 and Fermi motion and the
 important corrections that originate from $\Delta$ resonance modification 
inside the nuclear medium. Another issue is pion FSI for which we use a
simulation program that follows the work done in
 Ref.~\cite{Salcedo:1987md} where a general simulation code for inclusive
  pion nucleus
reactions was developed. In some of the channels  coherent pion
production  is also
possible and to evaluate its contribution we shall take our results in
Ref.~\cite{Hernandez:2010jf}
that uses the model we derived in Ref.~\cite{Amaro:2008hd}. 
\section{Pion production model}
\begin{figure}[tbh]
{\includegraphics[height=3.25cm]{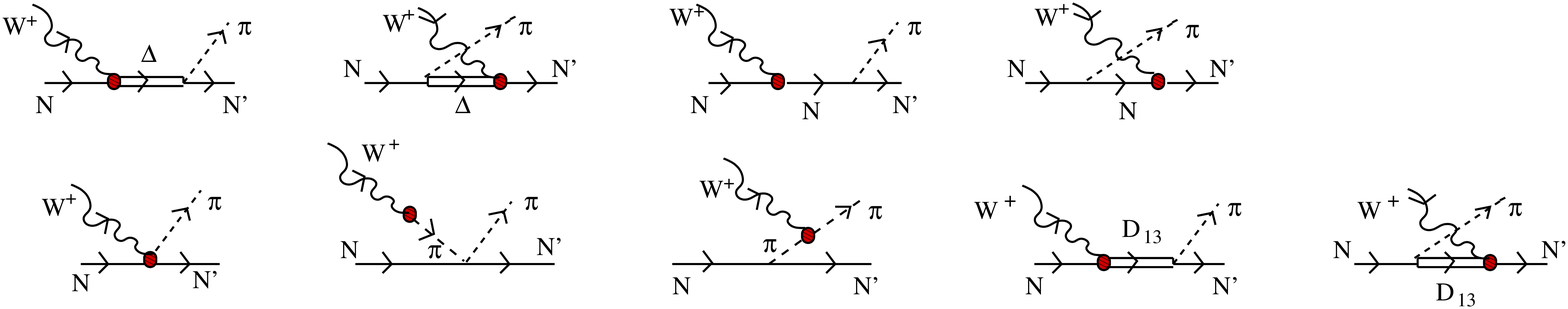}}
\caption{ Model for the $W^+N\to N^\prime\pi$
  reaction. We have Direct and crossed
  $\Delta(1232)-$  and nucleon  pole terms,
  contact and pion pole contribution, and  the
  pion-in-flight term~\cite{Hernandez:2007qq}.  In this case we also  include
    direct and crossed
  $D_{13}$-pole terms.}
  \label{fig:diagramas}
\end{figure}

\begin{figure}
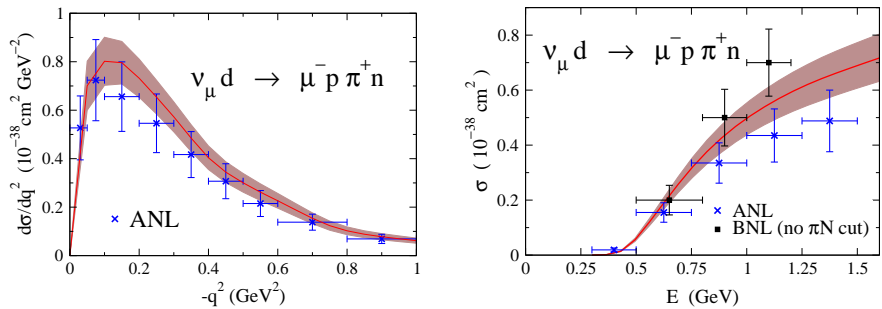

\makebox[0pt]{\hspace{1cm}\includegraphics[height=4cm]{nuint10_1.eps}\hspace{.5cm}
\includegraphics[height=4cm]{nuint10_2.eps}}
\label{fig:fit}
\caption{Comparison of our model results (solid line) to 
ANL~\cite{Radecky:1981fn}  and BNL~\cite{Kitagaki:1986ct}
experimental data. Theoretical 68\% confidence level bands are also shown. Data
include a systematic error (20\% for ANL and 10\% for BNL data) that has been
added in quadratures to the statistical published errors. Our theoretical
results and ANL data include a $W<1.4$\,GeV cut in the final $\pi N$ invariant mass. }
\end{figure}

Our full 
 model for one pion production on  the nucleon is depicted in 
 Fig.~\ref{fig:diagramas}. 
It contains the  dominant $\Delta$-pole resonance term (direct and crossed) and
background terms required by chiral symmetry. On top of that we add now the
direct and crossed
  $D_{13}$-pole terms. The background terms are the leading contributions
 of a $SU(2)$ nonlinear $\sigma$ model supplemented with well
   known form factors in a way that respects both conservation of vector 
   current
   and the partial conservation of axial current hypotheses.
 All the details on the $\Delta$ and background terms can be found in   
 Refs.~\cite{Hernandez:2007qq,Hernandez:2010bx}. In Ref.~\cite{Hernandez:2010bx}
  we followed the work in Ref.~\cite{Graczyk:2009qm} and  
 we made  a combined fit of the dominant nucleon-to-Delta axial form factor 
 to 
ANL and BNL data  including both full deuteron effects and  flux normalization
uncertainties. In Fig.~\ref{fig:fit} we show the results of that
fit compared to experimental data. The axial nucleon-to-Delta  form factors 
obtained in Ref.~\cite{Hernandez:2010bx} are the ones we use in the present 
calculation.  
As for the $D_{13}$ resonance contribution, all details will be given
elsewhere~\cite{ejm2013}. Note however that as the $D_{13}$ has isospin 1/2 it does not
contribute in the $p\pi^+$ channel and thus it does not affect the 
fit of the axial nucleon to Delta form factors carried out in Ref.~\cite{Hernandez:2010bx}.

 For incoherent production on a nucleus we  sum the nucleon cross section
 over all nucleons in the nucleus. For a charged current ($CC$) process, using
  the local density 
approximation, we arrive  
for initial pion production (prior to any pion FSI) induced by a neutrino of
momentum/energy $|\vec k|$  at
\begin{eqnarray*}
 \frac{d\sigma}{d k4\pi  r^2\,dr\,d\cos\theta_\pi\, dE_\pi}=\Phi(|\vec k |)\,
 \hspace{-.15cm}
  \sum_{N=n,p}
 2\int \frac{d^3p_N}{(2\pi)^3}\
\theta(E_F^N(r)-E_N) \,\theta(E_N+q^0-E_\pi-E_F^{N'}(r))
\frac{d\hat\sigma(\nu N\to l^-N'\pi)}{d\cos\theta_\pi dE_\pi}.\nonumber\\
\end{eqnarray*}
with
$E_F^N(r)=\sqrt{M^2+(k_F^N(r))^2}$, being $k_F^N(r)=(3\pi^2\rho_N(r))^{1/3}$
and $\rho_N(r)$ the local Fermi
momentum and  local density for nucleons of type $N$. Besides $\Phi(|\vec k|)$ is the neutrino flux.
$\hat\sigma(\nu N\to l^-N'\pi)$ is the cross section at the nucleon level modified
by medium effects as discussed below. The above differential
cross section is used in
a simulation code to generate, at a given point $\vec r$ inside the nucleus and by
neutrinos of a given energy, pions with a certain energy and momentum 
direction.

The $\Delta$ properties are strongly modified in the nuclear 
medium~\cite{Amaro:2008hd,Hirata:1978wp,Oset:1987re,Nieves:1993ev,Gil:1997bm,Benhar:2005dj,AlvarezRuso:2007tt,
AguilarArevalo:2010zc,Nieves:1991ye} and since the direct
$\Delta$-pole contribution is the dominant one a more
 correct treatment is needed for production inside a nucleus. Following Ref.~\cite{Gil:1997bm},
we
modify the  $\Delta$ propagator in the $\Delta$-pole term as
\begin{eqnarray*}
\frac1{p_\Delta^2-M_\Delta^2+iM_\Delta\Gamma_\Delta}\to\frac1{\sqrt{s}+M_\Delta}
\frac1{\sqrt{s}-M_\Delta+i(\Gamma^{\rm Pauli}_\Delta/2-{\rm Im}\Sigma_\Delta)},
\end{eqnarray*}
with $s=p_\Delta^2$, $\Gamma^{\rm Pauli}_\Delta$ the free $\Delta$ width
corrected by Pauli blocking of the final nucleon, for which we take the expression in Eq.(15) of
 Ref.~\cite{Nieves:1991ye}, and ${\rm Im}\Sigma_\Delta$
the imaginary part of the $\Delta$ self-energy in the medium. 
The evaluation of
$\Sigma_\Delta$ is done in Ref.~\cite{Oset:1987re} where the imaginary part is
parameterized as
\begin{eqnarray*}
-{\rm Im}\Sigma_\Delta=C_Q\left(\frac{\rho}{\rho_0}\right)^\alpha
+C_{A_2}\left(\frac{\rho}{\rho_0}\right)^\beta+
C_{A_3}\left(\frac{\rho}{\rho_0}\right)^\gamma,
\end{eqnarray*}
with $\rho_0=0.17\,$fm$^{-3}$. 
The terms in $C_{A_2}$ and $C_{A_3}$ are related to the two-body absorption
$WNN\to NN$ and three-body absorption $WNNN\to NNN$ channels respectively. On the other
hand the $C_Q$ term gives rise to a new $WN\to N\pi$ contribution   inside
the nuclear medium and thus it has to be taken into account beyond its role
in modifying 
the $\Delta$ propagator. This new contribution has to be added incoherently and
we implement it in a approximate way by taking as   amplitude square for 
this process the amplitude square of the
direct $\Delta$-pole contribution multiplied by 
 $\frac{C_Q(\rho/\rho_0)^\alpha}{\Gamma_\Delta/2}$.
When coherent production on $^{12}C$ is possible we evaluate its contribution
using our  model in
Ref.~\cite{Amaro:2008hd} but with the nucleon-to-Delta form factors as 
extracted in
  Ref~\cite{Hernandez:2010bx}. 

 As already mentioned, to evaluate  FSI effects   we follow  
Ref.~\cite{Salcedo:1987md} and we take into account
$P$- and $S$-wave pion absorption, and $P$-wave  quasielastic scattering 
on a single nucleon. The $P$- wave interaction
is mediated by the $\Delta$ resonance excitation. The different
contributions to the imaginary part of its self-energy account for
pion two- and three-nucleon absorption and quasielastic processes. The
probabilities for the different processes are evaluated in nuclear matter as a 
function of the density and then the local density approximation prescription
is used for its use in finite nuclei. After a quasielastic event, pions change
momentum and may change  its electric charge. The probability for charge
exchange and the final momentum distribution after a quasielastic interaction
are given in Ref.~\cite{Salcedo:1987md}. That information is used in the simulation
program to generate the pion resulting from such a collision. Besides, in between
collisions we assume the pions propagate in straight lines.  All the details
 can be found in Ref.~\cite{Salcedo:1987md}.
\section{Results and
comparison with MiniBooNE data}
\label{sec:results}
\begin{figure}[h!!]
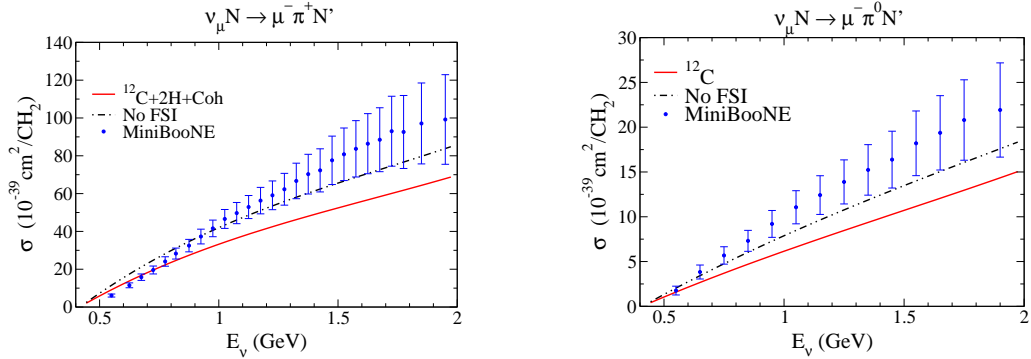

 \includegraphics[height=4.75cm]{2ccpipenergy.eps}\hspace{1.5cm}
\includegraphics[height=4.75cm]{1ccpi0energy.eps}
\caption{ $1\pi$ total production cross section 
for $\nu_\mu$ $CC$ interaction
in mineral oil. Left Panel: results for a final $\pi^+$. Right panel:
Results for a final $\pi^0$.  
Solid line: Total contribution. 
Double-dotted dashed line: Model prediction
without FSI of the outgoing pion. 
 Experimental data taken from
Ref.~\cite{AguilarArevalo:2010bm}. }
  \label{fig:totalccpip}
\end{figure}

Here we show part of the results we have obtained. A more complete discussion 
of the relevance of the different contributions will be given in Ref.~\cite{ejm2013}.
 In the left panel of Fig.~\ref{fig:totalccpip} we compare our results for 
 $\pi^+$ production in a $CC$ process with 
  MiniBooNE data. We take into account the
contribution on $^{12}C$ and  the two hydrogens. There is also a small
coherent contribution on  $^{12}C$. 
Our total result is below data for neutrino energies above $0.9$\,GeV. The
agreement improves if we do not take into account FSI of the pion.
A similar
result (see right panel of Fig.~\ref{fig:totalccpip}) is obtained for  a final $\pi^0$.\vspace{.1cm} 
\begin{figure}[tbh]
\includegraphics[height=4.75cm]{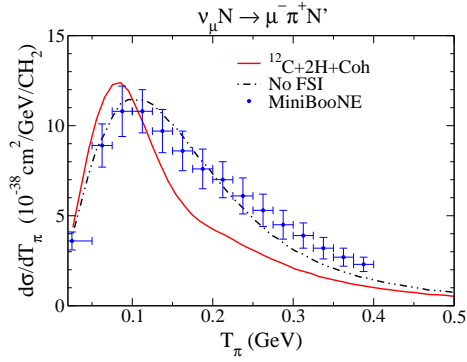}
\caption{ Differential $\frac{d\sigma}{dT_\pi}$ 
cross section for charged current $1\pi^+$ production by $\nu_\mu$ in mineral
oil. Captions as in Fig.~\ref{fig:totalccpip}. Data
 from Ref.~\cite{AguilarArevalo:2010bm}. }
  \label{fig:cctpi}
\end{figure}

\begin{figure}[tbh]
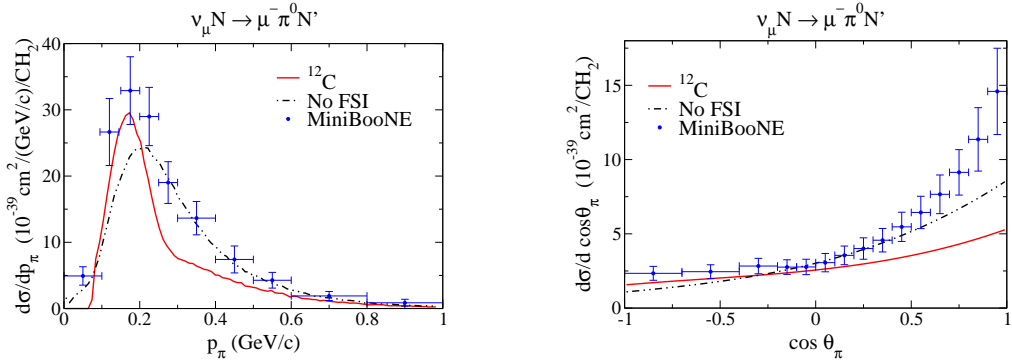

 \includegraphics[height=4.75cm]{1ccpi0mom.eps}\hspace{1.5cm}
\includegraphics[height=4.75cm]{1ccpi0cos.eps}
\caption{ Differential $\frac{d\sigma}{dp_\pi}$ (left panel) and
$\frac{d\sigma}{d\cos\theta_\pi}$ (right panel)
cross section for $CC$ $1\pi^0$ production by $\nu_\mu$ in mineral
oil. Captions as in Fig.~\ref{fig:totalccpip}. Data
 from Ref.~\cite{AguilarArevalo:2010xt}. }
  \label{fig:ccpi0}
\end{figure}

In Fig.~\ref{fig:cctpi} we compare the differential $\frac{d\sigma}{dT_\pi}$ 
cross section for $CC$ $1\pi^+$ production by $\nu_\mu$. 
We have taken into account the neutrino flux in  
 Ref.~\cite{AguilarArevalo:2010bm} to produce our results. 
 We underestimate data for
 $T_\pi$ above 0.15\,GeV. This is an effect of FSI of
 the final pion which above those kinetic energies accounts for a sizable
  pion absorption driven by the $\Delta (1232)$.
 Neglecting FSI we get a good agreement with data.

In Fig.\ref{fig:ccpi0} we show $\nu_\mu$ differential  $\frac{d\sigma}{dp_\pi}$ 
and  $\frac{d\sigma}{d\cos\theta_\pi}$ 
cross sections for $CC$ $1\pi^0$ production. For that we use the neutrino flux reported 
in Ref.~\cite{AguilarArevalo:2010xt} that extends from 2\,GeV down to 0.5\,GeV
neutrino energy.  Our results for $\frac{d\sigma}{dp_\pi}$ evaluated without FSI on the final pion agree better
with data for pion momentum above 0.2\,GeV/c. As a result of FSI, the agreement
improves below 0.2\,GeV/c, but our model produces too
few pions in the momentum region from 0.22 to 0.55\,GeV/c. The angular
distribution shows those missing pions mainly go in the forward direction.

\begin{figure}[tbh]
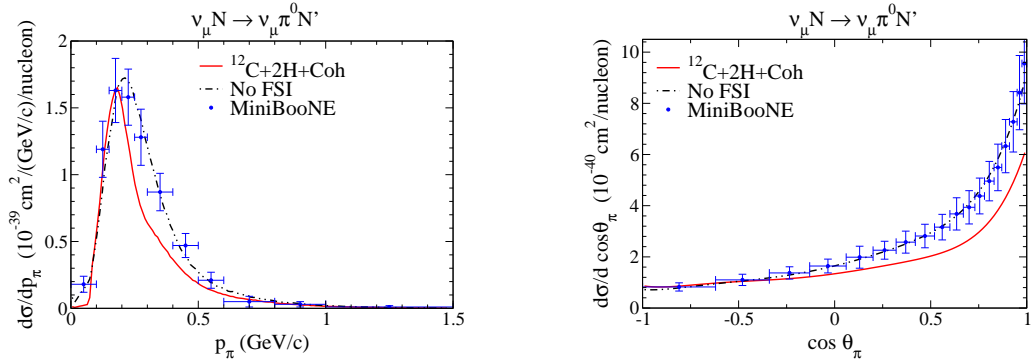

 \includegraphics[height=4.75cm]{3ncpi0mom.eps}\hspace{1.5cm}
\includegraphics[height=4.75cm]{3ncpi0cos.eps}
\caption{ Differential $\frac{d\sigma}{dp_\pi}$ (left panel) and 
$\frac{d\sigma}{d\cos\theta_\pi}$ (right panel) 
cross sections per nucleon for $NC$ $1\pi^0$ production by $\nu_\mu$ in mineral
oil. Captions as in Fig.~\ref{fig:totalccpip}. Data
 from Ref.~\cite{AguilarArevalo:2009ww}. }
  \label{fig:ncmomcos}
\end{figure}

In Fig.~\ref{fig:ncmomcos}  we show results for 
 $NC$ production induced by neutrinos that we compare with data by 
 the MiniBooNE Collaboration
in Ref.~\cite{AguilarArevalo:2009ww}. We use the 
$\nu_\mu$ flux reported by 
MiniBooNE. Our results for $\frac{d\sigma}{dp_\pi} $without FSI agree nicely
 with data,
  while our full model results show a depletion in the $0.25\sim0.5\,$GeV/c momentum 
region. The agreement with data is nevertheless better than in the $CC$ case.
 The differential $\frac{d\sigma}{d\cos\theta_\pi}$
cross section  is shown in the right panel of Fig.~\ref{fig:ncmomcos}. 
Once again our results without FSI
interaction of the final pion show a good agreement with experimental
 measurements. As for our full results, a clear deficit 
is seen in the
forward direction  but the agreement, as
it was the case for the $\frac{d\sigma}{dp_\pi}$ differential cross section, 
is better than in 
the corresponding $CC$ reaction. We obtain similar results for 
$NC$ production induced by antineutrinos, see Ref.~\cite{ejm2013}.  

Our results both for $CC$ and $NC$ processes are in good agreement 
with the  calculations in Refs.~\cite{Sobczyk.:2012zj,Lalakulich:2012cj}.
As it is the case there, we also find a better agreement with data if FSI is
ignored. The introduction of a formation time/zone, as done in 
Ref.~\cite{Golan:2012wx}, for pion production 
and its later interactions in the medium will decrease the effect of FSI and the agreement
with data will improve. On the other hand, in Ref.~\cite{Lalakulich:2012cj} it
is shown that the same FSI model applied to pion photoproduction on a nucleus
is able to give a fair reproduction of experimental data. In 
Ref.~\cite{ejm2013} we also show that our FSI model gives a fair reproduction of
pion photoproduction in nuclei so that it is not clear to us what are the 
cause for disagreement in the neutrino induced reactions.

\begin{theacknowledgments}
 This research was supported by  the Spanish Ministerio de Econom\'{\i}a y 
 Competitividad and European FEDER funds
under Contracts Nos. FPA2010-
21750-C02-02, FIS2011-28853-C02-01, FIS2011-28853-C02-02,  and the Spanish Consolider-Ingenio 2010
Programme CPAN (CSD2007-00042), by Generalitat
Valenciana under Contract No. PROMETEO/20090090
and by the EU HadronPhysics3 project, Grant Agreement
No. 283286.
\end{theacknowledgments}


\bibliographystyle{aipproc}

\end{document}